\documentclass[10pt,twocolumn]{revtex4}
\usepackage{slashed,graphicx}
\tighten

\begin{document}
\title{Spin polarized neutron matter           \\
within the Dirac-Brueckner-Hartree-Fock approach }
\author{F.~Sammarruca}
\address{Physics Department, University of Idaho, Moscow, ID 83844, U.S.A}
\author{P.~G.~Krastev}
\address{Physics Department, Texas A\&M University -- Commerce, Commerce, TX 75429-3011, U.S.A}
\date{\today}

\begin{abstract}
The relation between energy and density (known as the nuclear
equation of state) plays a major role in a variety of nuclear and
astrophysical systems. Spin and isospin asymmetries can have a
dramatic impact on the equation of state and possibly alter its
stability conditions. An example is the possible manifestation of
ferromagnetic instabilities, which would indicate the existence, at
a certain density, of a spin-polarized state with lower energy than
the unpolarized one. This issue is being discussed extensively in
the literature and the conclusions are presently very model
dependent. We will report and discuss our recent progress in the
study of spin-polarized neutron matter. The approach we take is
microscopic and relativistic. The calculated neutron matter
properties are derived from realistic nucleon-nucleon interactions.
This makes it possible to understand the properties of the equation 
of state in terms of specific features of the nuclear force model.

\end{abstract}
\maketitle

\renewcommand{\thesection}{\arabic{section}}
\section{Introduction}

The properties of dense and/or highly asymmetric nuclear matter,
where {\it asymmetric} may refer to isospin or spin asymmetries, are
of great current interest in nuclear physics and astrophysics.  This
topic is broad-scoped since it reaches out to exotic systems on the
nuclear chart  as well as, on a dramatically different scale,         
exotic objects in the universe such as compact stars.

In this paper, we investigate the bulk and single-particle
properties of spin-polarized neutron matter. The study of the
magnetic properties of dense matter is of considerable interest in
conjunction with the physics of pulsars, which are believed to be
rapidly rotating neutron stars with strong surface magnetic fields.
The polarizability of nuclear matter can have strong effects on
neutrino diffusion and, in turn, variations of  the neutrino mean
free path due to changes in the magnetic susceptibility of neutron
matter can impact the physics of supernovae and proton-neutron
stars.

The magnetic properties of neutron/nuclear matter have been studied
extensively since a long time by many authors and with a variety of
theoretical methods
\cite{pol1,pol2,pol3,pol4,pol5,pol6,pol7,pol8,pol9,pol10,pol11,pol25,pol12,
pol13,pol14,pol15,pol16,pol17,pol18,pol19,pol21,pol22,pol23,pol24,pol29,pol30,pol31,pol32,pol33}.
Nevertheless, conclusions about the possibility of a phase
transition to a ferromagnetic state at some critical density are
still contradictory. For instance, calculations based on Skyrme-type
interactions \cite{pol23} predict that such instabilities will occur
with increasing density. In particular, currently used Skyrme forces
show a ferromagnetic transition for neutron matter at densities
between 1.1$\rho_0$ and 3.5$\rho_0$ \cite{pol24}. On the other hand,
more recent predictions based on Monte Carlo simulations
\cite{pol15} and the Brueckner-Hartree-Fock (BHF) approach with
realistic nucleon-nucleon (NN) interactions \cite{pol17,pol18}
exclude these instabilities, at least at densities up to several
times normal nuclear density. Similarly, no evidence of a transition
to a ferromagnetic state was found in older calculations based on
the Brueckner-Hartree-Fock approach with the Reid hard-core
potential as well as a non-local separable potentials \cite{pol10}.
Relativistic calculations based on effective meson-nucleon
Lagrangians \cite{pol12} predict the ferromagnetic transition to
take place at several times nuclear matter density, with its onset
being crucially determined by the inclusion of the isovector mesons.
Clearly, the existence of such phase transition depends sensitively
on the modeling of the spin-dependent part of the nuclear force and
its behavior in the medium. Thus, this unsettled issue goes to the
very core of nuclear physics.

Our calculation is microscopic and treats the nucleons
relativistically. A parameter-free and internally consistent
approach is important if we are to interpret our conclusions in
terms of the underlying nuclear force. This is precisely our focus,
namely to understand the in-medium behavior of specific components
of the nuclear force (in this case, the spin dependence). Different
NN potentials can have comparable quality as seen from their global
description of NN data and yet differ in specific features. Thus, it
will be interesting to explore how, for a given many-body approach,
predictions for spin-polarized neutron matter depend upon specific
features of the NN potential. Second, it will be insightful to
compare with predictions based on a realistic NN potential and the
BHF method \cite{pol17}, especially at the higher densities, where
the repulsive Dirac effect can have a dramatic impact on the
short-range nature of the force.

This work is organized in the following way: after the introductory
notes in this section,  we briefly review our theoretical framework
(section 2); our results are presented and discussed in section 3;
we conclude in section 4 with a short summary and outlook.

\section{Brief description of the calculation}

The starting point of any microscopic calculation of nuclear
structure or reactions is a realistic free-space NN interaction. A
realistic and quantitative model for the nuclear force with
reasonable theoretical foundations is the one-boson-exchange (OBE)
model \cite{Machleidt89}. Unless otherwise specified, our standard
framework consists of the Bonn B potential together with the
Dirac-Brueckner-Hartree-Fock (DBHF) approach to nuclear matter. A
detailed description of our application of the DBHF method to
nuclear, neutron, and asymmetric matter can be found in  our earlier
works \cite{AS1,SBK,KS1}.

Similarly to what we have done to describe isospin asymmetries of
nuclear matter, the single-particle potential is the solution of a
set of coupled equations
\begin{equation}
U_u = U_{ud} + U_{uu}
\end{equation}
\begin{equation}
U_d = U_{du} + U_{dd}
\end{equation}
where $u$ and $d$ refer to ``up'' and ``down'' polarizations,
respectively, and where each $U_{\sigma \sigma '}$ term contains the
appropriate (spin-dependent) part of the interaction, $G_{\sigma
\sigma '}$. More specifically,
\begin{equation}
U_{\sigma}({\vec p}) = \sum _{\sigma '=u,d} \sum _{q\leq k_F^{\sigma
'}} <\sigma,\sigma '|G({\vec p},{\vec q})|\sigma,\sigma '>,
\end{equation}
where the second summation indicates integration over the two Fermi
seas of spin-up and spin-down neutrons, and
\begin{widetext}
\begin{eqnarray}
<\sigma,\sigma '|G({\vec p},{\vec q})|\sigma,\sigma '>&=&
\sum_{L,L',S,J,M,M_L} <\frac{1}{2} \sigma;\frac{1}{2} \sigma '|S
(\sigma + \sigma ')>
<\frac{1}{2} \sigma;\frac{1}{2} \sigma '|S (\sigma + \sigma ')> \nonumber\\
&\times&<L M_L;S(\sigma + \sigma ')|JM>
<L' M_L;S(\sigma + \sigma ')|JM> \nonumber\\
&\times& i^{L'-L} Y^{*}_{L',M_L}({\hat k_{rel}}) Y_{L,M_L}({\hat
k_{rel}}) <LSJ|G(k_{rel},K_{c.m.})|L'SJ>
\end{eqnarray}
\end{widetext}
The notation $<j_1m_1;j_2m_2|j_3m_3>$ is used for the Clebsh-Gordon
coefficients. Clearly, the need to separate the interaction by individual spin
components brings along angular dependence, with the result that the
single-particle potential depends also on the direction of the
momentum. Notice that the $G$-matrix equation is solved using
partial wave decomposition and the matrix elements are then summed
as in Eq.~(4) to provide the new matrix elements in the
uncoupled-spin representation needed for Eq.~(3). The
three-dimensional integration in Eq.~(3) is performed in terms of
the spherical coordinates of ${\vec q}$, $(q,\theta_q,\phi_q)$, with
the final result depending upon both magnitude and direction of
${\vec p}$. On the other hand, the scattering equation is solved
using relative and center-of-mass coordinates, $k_{rel}$ and
$K_{c.m.}$. These are easily related to the momenta of the two
particles in the nuclear matter rest frame through the standard
definitions ${\vec K}_{c.m.}={\vec p}+{\vec q}$, and ${\vec
k}_{rel}=\frac{{\vec p}-{\vec q}}{2}$. (The latter displays the dependence
of the argument of the spherical harmonics upon ${\vec p}$ and
${\vec q}$.) 

Solving the $G$-matrix equation requires knowledge of the
single-particle potential, which in turn requires knowledge of the
interaction. Hence, Eqs.~(1-2) together with the $G$-matrix equation
constitute a self-consistency problem, which is handled,
technically, exactly the same way as previously done for the case of
isospin asymmetry \cite{AS1}. The Pauli operator for scattering of
two particles with unequal Fermi momenta, contained in the kernel of
the $G$-matrix equation, is also defined in perfect analogy with the
isospin-asymmetric one \cite{AS1},
\begin{equation}
Q_{\sigma \sigma '}(p,q,k_F^{\sigma},k_F^{\sigma '})=\left\{
\begin{array}{l l}
1 & \quad \mbox{if $p>k_F^{\sigma}$ and  $q>k_F^{\sigma '}$}\\
0 & \quad \mbox{otherwise.}
\end{array}
\right.
\end{equation}
Notice that, although a full three-dimensional integration is
performed in Eq.~(3), the usual angle-average procedure is applied
to the Pauli operator (when expressed in terms of $k_{rel}$ and
$K_{c.m.}$) and to the two-particle propagator in the kernel of the
$G$-matrix equation.

Once a self-consistent solution is obtained for the single-particle
spectrum, the average potential energy for particles with spin
polarization $\sigma$ is obtained as
\begin{equation}
<U_{\sigma}> = \frac{1}{2} \frac{1}{(2 \pi)^3}
\frac{1}{\rho_{\sigma}} \int _{p=0}^{k_F^{\sigma}} \int
_{\theta_p=0}^{\pi} \int _{\phi_p=0}^{2 \pi} U_{\sigma}({\vec
p})p^2dp d \Omega _p
\end{equation}
The average potential energy per particle is then
\begin{equation}
<U> = \frac{\rho_{u}<U_{u}> + \rho_{d}<U_{d}>}{\rho}
\end{equation}

The kinetic energy (or, rather, the free-particle operator in the
Dirac equation when using the DBHF framework), is also averaged over
magnitude and direction of the momentum. In particular, we calculate
the average free-particle energy for spin-up(down) neutrons as
\begin{equation}
<T_{\sigma}> = \frac {\int d \Omega \bar{T}(m^*_{\sigma}(\theta))}
{\int d \Omega}
\end{equation}
where $\bar{T}$ is the average over the magnitude of the momentum.
Notice that the angular dependence comes in through the effective
masses, which, being part of the parametrization of the
single-particle potential, are themselves direction dependent (and
of course different for spin-up or spin-down neutrons).

Finally
\begin{equation}
<T> = \frac{\rho_{u}<T_{u}> + \rho_{d}<T_{d}>}{\rho}
\end{equation}
and the average energy per neutron is
\begin{equation}
\bar{e} = <U> + <T>
\end{equation}

As in the case of isospin asymmetry, it can be expected that the
dependence of the average energy per particle upon the degree of
polarization \cite{pol17} will follow the law
\begin{equation}
\bar{e}(\rho, \beta) = \bar{e}(\rho, \beta =0) + S(\rho) \beta ^2
\end{equation}
where $\beta$ is the spin asymmetry, defined by
$\beta=\frac{\rho_{u}-\rho_{d}}{\rho}$. A negative value of
$S(\rho)$ would signify that a polarized system is more
stable than unpolarized neutron matter.

From the energy shift,
\begin{equation}
S(\rho)=\bar{e}(\rho,\beta=1)-\bar{e}(\rho,\beta=0),
\end{equation}
the magnetic susceptibility can be easily calculated. If the
parabolic dependence is assumed, then one can write the magnetic susceptibility  as \cite{pol17}
\begin{equation}
\chi =\frac{ \mu^2 \rho}{2S(\rho)}, 
\end{equation}
where $\mu$ is the neutron magnetic moment. The magnetic
susceptibility in often expressed in units of $\chi _F$, the
magnetic susceptibility of a free Fermi gas,
\begin{equation}
\chi _F =\frac{ \mu^2 m}{\hbar ^2 \pi ^2}k_F ,
\end{equation}
where $k_F$ denotes the average Fermi momentum which is related to
the total density by
\begin{equation}
k_F = (3\pi ^2 \rho)^{1/3} . 
\end{equation}
The Fermi momenta for up and down neutrons are
\begin{eqnarray}
k_F^u = k_F(1 + \beta)^{1/3} \nonumber\\
k_F^d = k_F(1 -\beta)^{1/3}. 
\end{eqnarray}

\begin{figure}[!t]
\centering
\includegraphics[totalheight=3.0in]{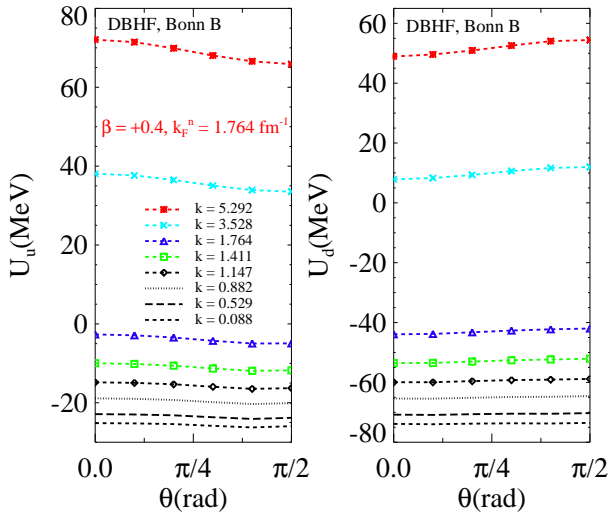}
\caption{(Color online) Angular dependence of the single-particle
potential for spin-up and spin-down neutrons at fixed spin asymmetry
and Fermi momentum and for different values of the neutron momentum.
The momenta are in units of $fm^{-1}$. The angle is defined relative
to the polarization axis.}
\end{figure}

\begin{figure}[!t]
\centering
\includegraphics[totalheight=3.0in]{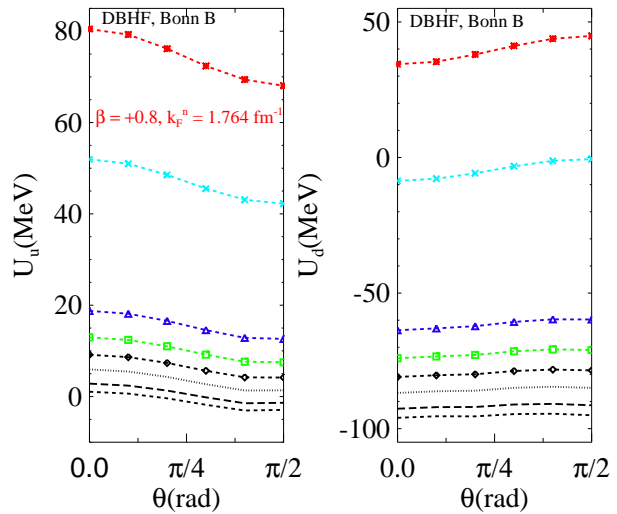}
\caption{(Color online) As in the previous figure but for a larger
value of the spin asymmetry. }
\end{figure}
\begin{figure}[!b]
\centering
\includegraphics[totalheight=2.25in]{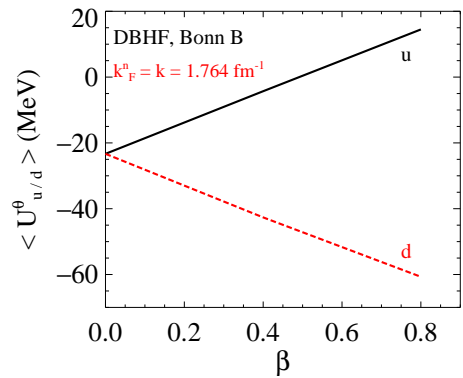}
\caption{(Color online) Asymmetry dependence of the single-particle
potential for spin-up and spin-down neutrons at fixed density and
momentum. The angular dependence is integrated out. }
\end{figure}
\begin{figure}[!t]
\centering
\includegraphics[totalheight=2.25in]{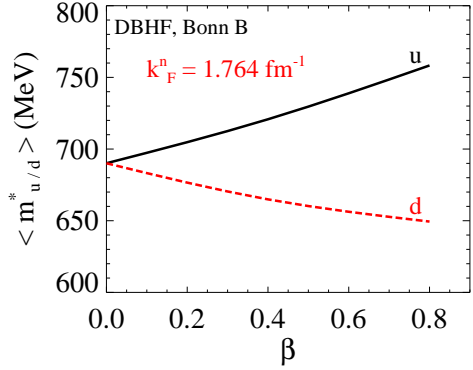}
\caption{(Color online) Asymmetry dependence of the effective masses
for upward and downward polarized neutrons under the same conditions
as in Fig.~3. }
\end{figure}
\begin{figure}[!t]
\centering
\includegraphics[totalheight=3.5in]{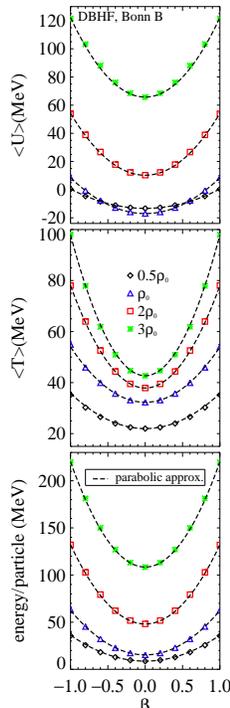}
\caption{(Color online) Average potential, kinetic, and total energy
per particle at various densities as a function of the spin
asymmetry. Predictions obtained with our standard DBHF calculation.}
\end{figure}
\begin{figure}[!t]
\centering
\includegraphics[totalheight=3.5in]{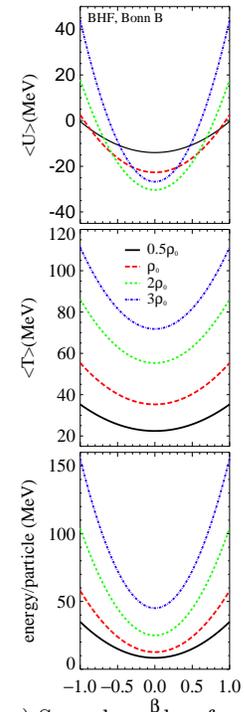}
\caption{(Color online) Same legend as for the previous figure, but
the predictions are obtained with the BHF calculation. }
\end{figure}
\begin{figure}[!t]
\centering
\includegraphics[totalheight=2.25in]{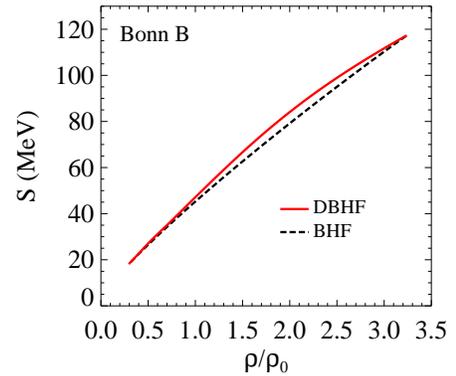}
\caption{(Color online) Energy difference between the polarized and the 
unpolarized states corresponding to Fig.~5 and 6.}
\end{figure}
\begin{figure}[!t]
\centering
\includegraphics[totalheight=2.25in]{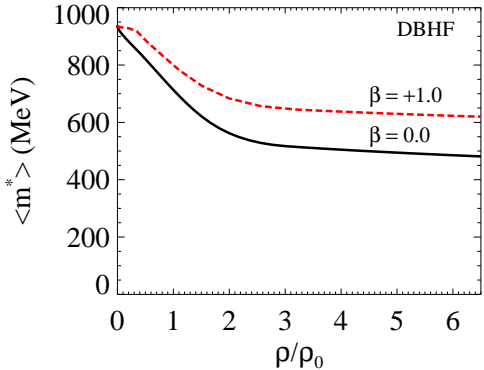}
\caption{(Color online) Neutron effective masses used in the DBHF    
calculations of the EOS. The angular dependence is averaged out.}
\end{figure}

\begin{figure}[!t]
\centering
\includegraphics[totalheight=3.0in]{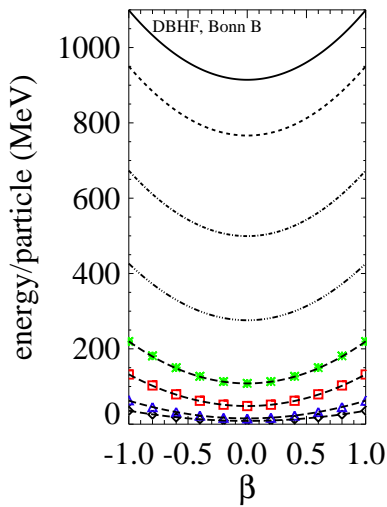}
\caption{(Color online) Average energy
per particle at densities equal to 0.5,1,2,3,5,7,9, and
10 times $\rho_0$ (from lowest to highest curve). 
Predictions obtained with our DBHF calculation.}
\end{figure}
\begin{figure}[!b]
\centering
\includegraphics[totalheight=2.25in]{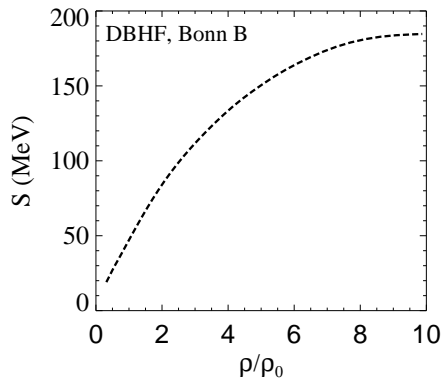}
\caption{(Color online) Density dependence of the spin symmetry
energy obtained with the DBHF model.}                                      
\end{figure}

\begin{figure}[!h]
\centering
\includegraphics[totalheight=2.25in]{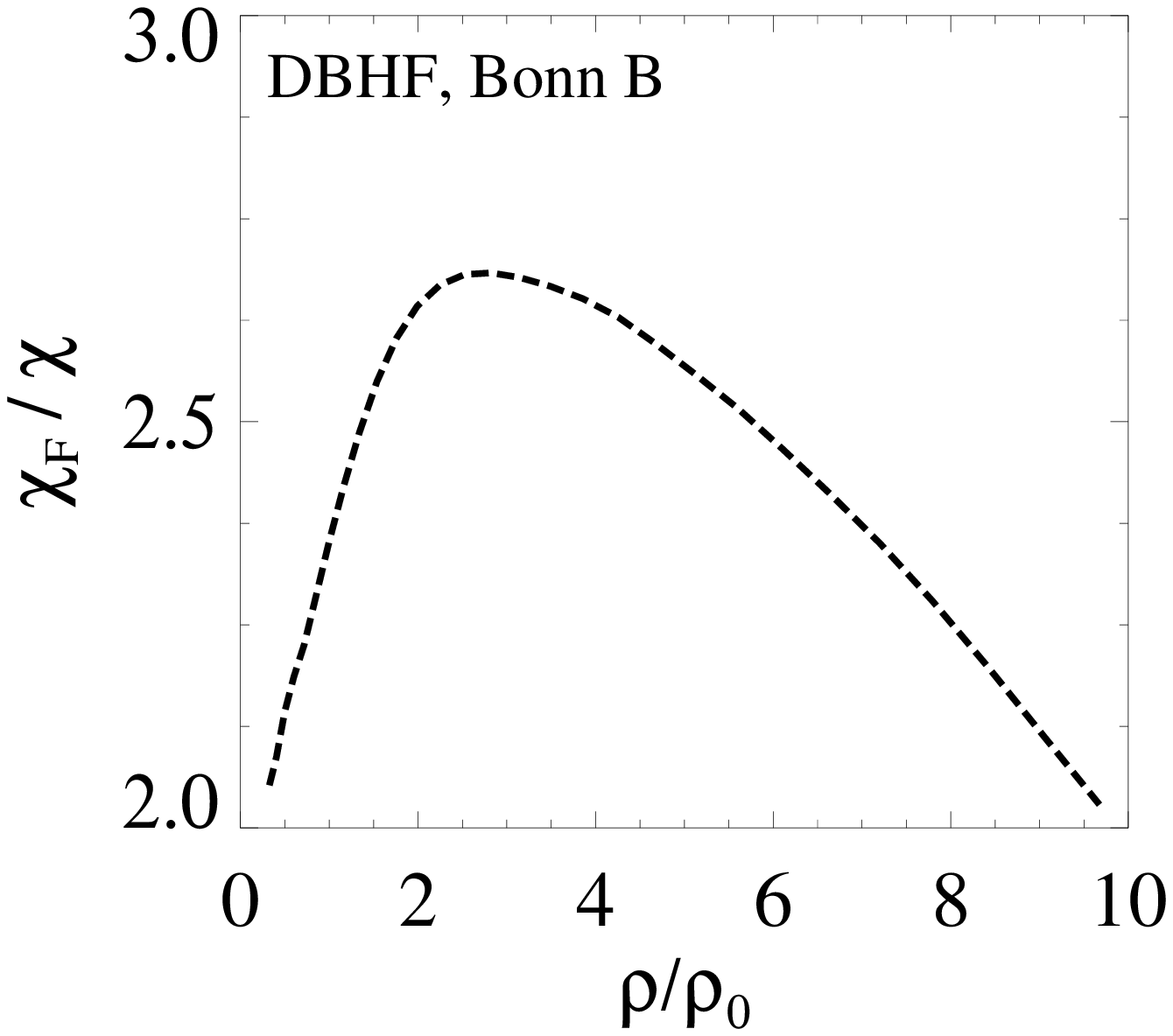}
\caption{(Color online) Density dependence of the ratio $\chi_F
/\chi $. Predictions are obtained with the DBHF model.       
} 
\end{figure}

For the most general case, it will be necessary to combine isospin
and spin asymmetry. With twice as many degrees of freedom, the
coupled self-consistency problem schematically displayed in
Eqs.(1-2) is numerically more involved but straightforward. This is
left to a future work.      

\section{Results and discussion}

We begin by showing the angular and momentum dependence of the
single-neutron potential, see Figs.~1-2. The angular dependence is
rather mild, especially at the lowest momenta. As can be reasonably
expected, it becomes stronger at larger values of the asymmetry, see
Fig.~2. In Fig.~3, the asymmetry dependence is displayed for fixed
density and momentum (here the angular dependence is averaged out).
As the density of $u$ particles goes up, the total density remaining
constant, the most likely kind of interaction for $u$ neutrons is of
the $uu$ type. Similarly, the largest contribution to the
$d$-particle potential is of the $du$ type, see Eqs.~(1-2), with the
latter being apparently more attractive, as can be inferred by the
spin splitting of the potential shown in Fig.~3. Before we move on
to discuss this issue in greater details, we also show the effective
masses of $u$ and $d$ neutrons, see Fig.~4, and observe that they
display a qualitatively similar behavior as the one of the
corresponding single-particle potentials.

The average energy per particle at various densities and as a
function of the asymmetry parameter is shown in the third frame of
Fig.~5. The first two frames display the contribution from the
average potential energy and the average kinetic energy,
respectively. The parabolic dependence on $\beta$, or linear on
$\beta ^2$, is obviously verified. In Fig.~6 we show the
corresponding predictions obtained with the conventional
Brueckner-Harteee-Fock approach. This comparison may be quite
insightful, as we further discuss next. We notice that the Dirac
energies are overall more repulsive, but the parabolas predicted
with the BHF prescription appear to become steeper, relative to each 
other, as density grows. 
The energy difference between the totally
polarized state and the unpolarized one for both the relativistic and the 
non-relativistic calculation is shown in Fig.~7. Although initially 
higher, the growth of the DBHF curve shows a tendency to slow down and the two sets
of predictions cross over just above 3$\rho_0$.

Before leaving this detour into the non-relativistic model, we observe that the predictions shown in Fig.~6
are reasonably consistent with those from 
previous studies which used
the Brueckner-Hartree-Fock approach and the Nijmegen II and Reid93
NN potentials \cite{pol17}. In fact, comparison with that work
allows us to make some useful observations concerning the choice of
a particular NN potential, for a similar many-body approach (in this
case, BHF). We must keep in mind that off-shell differences exist
among NN potentials (even if nearly equivalent in their fit of NN
scattering data) and those will impact the $G$-matrix (which, unlike
the $T$-matrix, is not constrained by the two-body data).
Furthermore, off-shell differences will have a larger impact at high
Fermi momenta, where the higher momentum components of the NN
potential, (usually also the most model dependent), play a larger
role in the calculation. Accordingly, the best agreement between our
BHF predictions and those of Ref.~\cite{pol17} is seen at low to
moderate densities.
Furthermore, as far as differences based on the choice of the NN
potential are concerned, we would expect them to be more pronounced
for nuclear matter than for pure neutron matter, since the largest
variations among modern realistic potentials are typically found in
the strength of the tensor force, which is stronger in T=0 partial
waves (obviously absent in the $nn$ system). This point will be
explored in a later investigation.

In the remainder of this paper, we will focus on the DBHF model, which
is our standard operational approach. To further explore the possibility
of a ferromagnetic transition, 
we have extended the DBHF calculation to 
densities as high as 10$\rho_0$.                                                 
The same method as described in Ref.~\cite{KS1} is applied to obtain the
energy per particle                           
where a self-consistent solution cannot be obtained 
(see Section III of Ref.~\cite{KS1} for details). The (angle-averaged)
neutron effective masses for both the unpolarized and the fully polarized case
are shown in Fig.~8 as a function of density.

DBHF predictions for 
the average energy per particle are shown in Fig.~9 at densities ranging from   
$\rho$=0.5$\rho_0$ to  
10$\rho_0$.    
What we observe is best seen through 
the spin-symmetry energy, which
we calculate from Eq.~(12) and show in Fig.~10.                            
We see that at high density the energy shift between
polarized and unpolarized matter continues to grow, but at a smaller 
rate, and eventually appear to saturate.
Similar observations already made in
conjunction with isospin asymmetry were explained in terms of
stronger short-range repulsion in the Dirac model \cite{KS1}. It
must be kept in mind that some large contributions, such as the one
from the $^1S_0$ state, are not allowed in the fully polarized case.
Now, if such contributions (typically attractive at normal
densities) become more and more repulsive with density (due to the
increasing importance of short-range repulsive effects), their
absence will amount to less repulsive energies at high density. On
the other hand, if large and attractive singlet partial waves remain
attractive up to high densities, their suppression (demanded in the
totally polarized case) will effectively amount to increased
repulsion.

In conclusion, although the curvature of the spin-symmetry energy may suggest
that ferromagnetic instabilities are in           
principle possible within the Dirac model, inspection of
Fig.~10 reveals that such transition does not take place at least  
up to 10$\rho_0$. Clearly 
it would not be appropriate to explore even higher densities without 
additional considerations, such as transition to a quark phase.
In fact, even on the high side of the densities considered here,
softening of the
equation of state from additional degrees of freedom not included in
the present model may be necessary in order to draw a more definite
conclusion.

Finally, in Fig.~11 we show the ratio $\chi _F / \chi$, whose
behavior is directly related to the spin-symmetry energy, see
Eq.~(13). Clearly, similar observations apply to both Fig.~11 and
Fig.~10. (The magnetic susceptibility would show an infinite
discontinuity, corresponding to a sign change of $S(\rho)$, in case
of a ferromagnetic instability.)

\section{Conclusions}
We have calculated bulk and single-particle properties of
spin-polarized neutron matter. The EOSs we obtain with the DBHF
model are generally rather repulsive at the larger densities. The
energy of the unpolarized system (where all $nn$ partial waves are
allowed), grows rapidly at high density with the result that the
energy difference between totally polarized and unpolarized neutron
matter tends to slow down with density.      
This may be interpreted as a {\it precursor} of
spin-separation instabilities, although no such transition is actually seen 
up to 10$\rho_0$. 
Our analysis allowed us to                  
locate the origin of this behavior 
in the contributions to the energy from specific partial
waves and their behavior in the medium, particularly increased repulsion
in the singlet states.

In future work, the impact of further extensions will be considered,
such as: examining the effects of contributions that soften the
EOS (especially at high density); extending our framework to
incorporate both spin- and isospin-asymmetries; examining the
temperature dependence of our observations for spin- and
isospin-asymmetries of neutron and nuclear matter.

\section*{Acknowledgements}

\noindent The authors acknowledge financial support from the U.S.
Department of Energy under grant number DE-FG02-03ER41270.

\end{document}